\documentclass[aps,prb,10pt,twocolumn,english,superscriptaddress,citeautoscript,showkeys,preprintnumbers,amsmath,amssymb,floatfix,footinbib]{revtex4-2}
\usepackage[colorlinks=true,linkcolor=blue ,citecolor=blue,urlcolor=blue]{hyperref}
\usepackage{amsmath}
\usepackage{graphicx}
\usepackage{dcolumn}
\usepackage[dvipsnames]{xcolor}
\usepackage{soul}
\usepackage{adjustbox}
\usepackage{lipsum}
\usepackage{blindtext}
\usepackage{lipsum}

\mathchardef\mhyphen="2D

\begin{document}

\title{Hidden Magnetic Complexity Within a Simple van der Waals Ferromagnet Ce$_2$Te$_5$}

\author{Rajesh Tripathi}
\email{raj7tpi@gmail.com}
\affiliation{ISIS Neutron and Muon Source, STFC, Rutherford Appleton Laboratory, Chilton, Oxon OX11 0QX, United Kingdom}
\affiliation{Jawaharlal Nehru Centre for Advanced Scientific Research, Jakkur, Bangalore 560064, India}
\affiliation{Department of Physics, Indian Institute of Technology (BHU) Varanasi, 221005 Uttar Pradesh, India}

\author{D. T. Adroja}
\email{devashibhai.adroja@stfc.ac.uk}
\affiliation{ISIS Neutron and Muon Source, STFC, Rutherford Appleton Laboratory, Chilton, Oxon OX11 0QX, United Kingdom}
\affiliation{Highly Correlated Matter Research Group, Physics Department, University of Johannesburg, Auckland Park 2006, South Africa}

\author{C. Ritter}
\email{ritter@ill.fr}
\affiliation{Institut Laue-Langevin, 71 Avenue des Martyrs, CS 20156, 38042, Grenoble Cedex 9, France}

\author{Leandro Liborio}
\affiliation{Scientific Computing Department, STFC, Rutherford Appleton Laboratory, Harwell Campus, Didcot OX11 0QX, United Kingdom}

\author{Manuel dos Santos Dias}
\affiliation{Scientific Computing Department, STFC Daresbury Laboratory, Sci-Tech Daresbury, Keckwick Lane, Daresbury WA4 4AD, United Kingdom}

\author{Leon Petit}
\affiliation{Scientific Computing Department, STFC Daresbury Laboratory, Sci-Tech Daresbury, Keckwick Lane, Daresbury WA4 4AD, United Kingdom}

\author{D. Khalyavin}
\affiliation{ISIS Neutron and Muon Source, STFC, Rutherford Appleton Laboratory, Chilton, Oxon OX11 0QX, United Kingdom}

\author{Yu Liu}
\affiliation{MPA-Q, Los Alamos National Laboratory, Los Alamos, New Mexico 87545, USA}

\author{T. Shiroka}
\affiliation{PSI Center for Neutron and Muon Sciences, Paul Scherrer Institut, CH-5232 Villigen PSI, Switzerland}

\author{M. Aouane}
\affiliation{ISIS Neutron and Muon Source, STFC, Rutherford Appleton Laboratory, Chilton, Oxon OX11 0QX, United Kingdom}
\affiliation{European Spallation Source ERIC,SE-221 00, Lund, Sweden}
\author{S. Langridge}
\affiliation{ISIS Neutron and Muon Source, STFC, Rutherford Appleton Laboratory, Chilton, Oxon OX11 0QX, United Kingdom}

\author {S. Patil}
\affiliation{Department of Physics, Indian Institute of Technology (BHU) Varanasi, 221005 Uttar Pradesh, India}

\author{Eric D. Bauer}
\affiliation{MPA-Q, Los Alamos National Laboratory, Los Alamos, New Mexico 87545, USA}

\date{\today}

\begin{abstract}
Ce$_2$Te$_5$ is a layered $f$-electron van der Waals magnet in which reduced dimensionality and inequivalent Ce sites give rise to competing magnetic interactions. We investigate its magnetic ground state using muon spin relaxation ($\mu$SR), neutron powder diffraction (NPD), and inelastic neutron scattering (INS). Zero-field $\mu$SR reveals an onset of static magnetism below $T_{\mathrm{C}} = 5.0(1)$~K, followed by an additional anomaly in the internal field at $T_{\mathrm{2}} = 2.3(2)$~K, consistent with features observed in bulk thermodynamic and transport measurements. In contrast, NPD data collected between $0.05-8$~K  reveal a single long-range ordered magnetic phase below $T_{\mathrm C}$, with the magnetic Bragg intensities vanishing at $T_{\mathrm C}$ with no evidence for additional structural or magnetic phase transitions down to base temperature. The ordered state is characterized by a commensurate propagation vector $\mathbf{k}=(0,0,0)$ and ferromagnetic alignment of Ce moments along the crystallographic $b$ axis. Remarkably, only one of the two crystallographically distinct Ce sites carries an ordered moment of $\sim 0.85(3)~\mu_{\mathrm B}$ per Ce at 0.05~K. INS measurements establish the crystal electric field (CEF) energy scale of Ce$^{3+}$, revealing low-lying excitations at 7.94 and 21.46~meV and a Kramers doublet ground state with strong single-ion anisotropy. These results demonstrate that Ce$_2$Te$_5$ undergoes a single symmetry-breaking magnetic transition, while additional low-temperature anomalies reflect subtle modifications of the ordered state driven by competing interactions and CEF effects.


\end{abstract}

\maketitle
\section{INTRODUCTION}
\label{Intro}

Reduced dimensionality often enhances the role of fluctuations, competing interactions, and magnetic anisotropy, giving rise to collective phenomena that are absent in three-dimensional materials. Layered van der Waals (vdW) materials provide an ideal platform for exploring such effects because their weak interlayer bonding enables the realization of atomically thin crystals, heterostructures, and twisted multilayers with highly tunable electronic and magnetic properties~\cite{Geim2013,Novoselov2016,Mannix2017}. Recent developments in magnetic vdW materials have further expanded this field through the discovery of intrinsic magnetic order in atomically thin crystals. Representative examples include the insulating ferromagnets (FM) CrI$_3$~\cite{huang2017layer} and Cr$_2$Ge$_2$Te$_6$~\cite{gong2017discovery}, the metallic FM Fe$_3$GeTe$_2$, with gate-tunable magnetic properties~\cite{Deng2018}, and the ultrathin CrTe$_2$, known to exhibit intrinsic FM up to room temperature~\cite{Zhang2021}. These breakthroughs have established low-dimensional magnetism as a versatile platform for exploring fundamental spin phenomena and for developing next-generation spintronic and quantum technologies~\cite{Burch2018}.

In contrast to transition-metal vdW magnets, comparatively little is known about layered $f$-electron systems such as CeSiI~\cite{PhysRevMaterials.5.L121401}, EuC$_6$~\cite{C9MH01988J,Lamura2012}, and GdTe$_3$~\cite{doi:10.1126/sciadv.aay6407}. These materials introduce additional degrees of freedom through strong spin--orbit coupling, crystal electric field (CEF) effects, and the interaction between localized $4f$ moments and itinerant conduction electrons. The competition among these interactions can produce pronounced magnetic anisotropy, multiple magnetic energy scales, and complex phase diagrams with successive ordering phenomena. Combined with reduced dimensionality, these effects provide a fertile environment for unconventional magnetic behavior and correlated ground states that are difficult to realize in conventional $3d$-electron magnets. Consequently, rare-earth vdW materials have emerged as a promising platform for exploring the interplay between low-dimensionality, competing magnetic interactions, and strong electronic correlations.

In this context, the family of rare-earth telluride $R$Te$_x$ ($R$ = La-Nd, Sm, and Gd-Tm; $x$ = 2, 2.5, and 3) has garnered significant interest owing to their incommensurate charge density wave (CDW)~\cite{PhysRevB.52.14516,PhysRevB.78.045123,malliakas2006divergence}, rich magnetic properties~\cite{PhysRevB.78.012410,shin2000anisotropic}, and the onset of superconductivity under pressure ($R$ = Gd, Tb, and Dy)~\cite{PhysRevB.91.205114}.

\begin{figure}
		\includegraphics[width=5 cm]{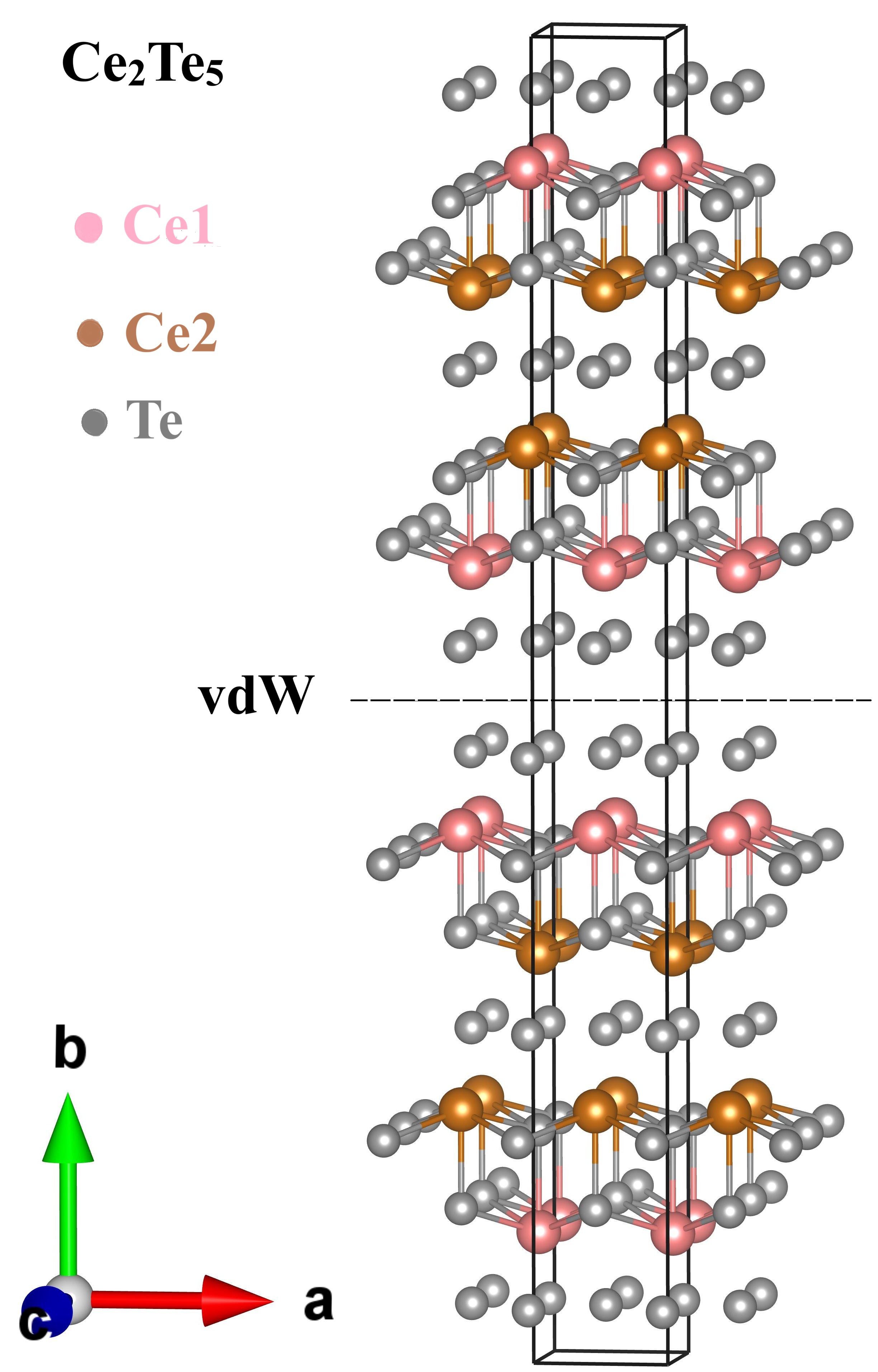}
		\caption{Unit-cell representation of Ce$_2$Te$_5$ (space group $Cmcm$), illustrating the two inequivalent Ce sites (Ce1 and Ce2) embedded within the Te framework. The structure consists of layers stacked along the $b$ axis, with vdW gaps separating adjacent Ce-Te slabs. The crystallographic axes are indicated.}
		\label{unit_cell}
\end{figure}

$R$Te$_x$ compounds possess a layered crystal structure formed by double Te square-net sheets (perpendicular to the $c$ axis), separated by double-corrugated RTe slabs, as illustrated in Fig.~\ref{unit_cell}. Among this series, CeTe$_2$ and CeTe$_3$ crystallize in the layered Cu$_2$Sb-type tetragonal structure (space group $P4/nmm$) and the weakly orthorhombic NdTe$_3$-type structure (space group $Cmcm$), respectively. Both structures feature localized Ce$^{3+}$ magnetic moments. CeTe$_2$ contains a single Te layer and undergoes an antiferromagnetic (AFM) transition at $T_\mathrm{N} = 4.3$\,K~\cite{KWON19951173,MIN2002203,doi:10.1143/JPSJ.69.937,PhysRevB.62.11609}. At 2~K, a metamagnetic transition occurs at a low magnetic field of 0.06~T, leading to a field-induced FM state with an easy $c$-axis orientation. Neutron diffraction studies reveal a down-up-up-down AFM configuration along the $c$-axis, with FM Ce double layers above and below the Te layer in CeTe$_2$~\cite{PARK1997684,PARK2000451,STOWE2000101,LIU2003151,10.1063/1.1853831}. On the other hand, CeTe$_3$ contains two neighboring Te square-net sheets connected via a weak vdW interaction. It exhibits two AFM transitions at $T_{N1}$ = 3.1~K and $T_{N2}$ = 1.3~K, with non-parallel easy axes perpendicular to the layer stacking direction, indicating a strongly easy-plane character~\cite{PhysRevB.79.134428,KDeguchi2009,10.1063/5.0007517}.
Ce$_2$Te$_5$ can be regarded as a combination of CeTe$_2$ and CeTe$_3$. It features alternating single and double Te layers stacked along the $b$-axis of the orthorhombic unit cell, separated by Ce-Te slabs~\cite{Chen2017}. The crystal structure of Ce$_2$Te$_5$ resembles that of CeTe$_3$, with two distinct Ce sites: Ce1 adjacent to the double layers of Te and Ce2 adjacent to the monolayer of Te. Chen et al. reported three magnetic transitions at 5.1~K, 2.3~K, and 0.9~K in single crystals of Ce$_2$Te$_5$~\cite{Chen2017}, while Liu et al. reported four magnetic transitions at 5.2~K, 2.1~K, 0.9~K, and 0.4~K in single crystals of Ce$_2$Te$_5$~\cite{PhysRevMaterials.6.094407}.

In this work, we address the microscopic origin of the multiple low-temperature anomalies reported in Ce$_2$Te$_5$ by combining muon spin relaxation ($\mu$SR), neutron powder diffraction (NPD), and inelastic neutron scattering (INS) measurements. We find that Ce$_2$Te$_5$ develops a single long-range FM state below $T_{\mathrm C}$ characterized by a propagation vector $\mathbf{k}=(0,0,0)$ and moments aligned along the crystallographic $b$ axis. Remarkably, only one of the two crystallographically inequivalent Ce sites carries a sizable ordered moment. Furthermore, crystal electric field (CEF) excitations at approximately 8 and 21~meV establish a strongly anisotropic Ce$^{3+}$ ground state. These results demonstrate that the additional low-temperature anomalies do not arise from distinct magnetic phases but instead reflect more subtle modifications of the ordered state driven by competing interactions and CEF effects.

\section{EXPERIMENTAL METHODS}
\label{expt}
Single crystals of Ce$_2$Te$_{5}$ were prepared following the procedure reported in Ref.~\cite{PhysRevMaterials.6.094407}. Zero-field (ZF) $\mu$SR experiments were performed on crushed single crystals of Ce$_2$Te$_{5}$ using a He-3 cryostat at the Dolly spectrometer of the Swiss Muon Source (S$\mu$S), Paul Scherrer Institute (PSI), Villigen, Switzerland. We used a 13-mm diameter $\times$ 2-mm thick powder pellet of Ce$_2$Te$_5$, which was mounted on the sample stick. The temperature range of the measurements was from 0.27 to 5.4~K, allowing investigations of the multiple magnetic ordering transitions and low-temperature magnetic ground state.

The muon stopping sites were calculated using the unperturbed electrostatic potential (UEP) method, as implemented in the software package pymuon-suite and in the Galaxy workflow management platform~\cite{Liborio2018,Sturniolo2019,Sturniolo2020,Sturniolo2023}.
The UEP method uses density functional theory (DFT) calculations to estimate the host material electrostatic potential, plus a combination of mathematical analysis and clustering techniques to estimate potential muon stopping sites.
The DFT simulations were performed with the CASTEP code~\cite{Clark2005}, using a spin polarization, a plane wave cut-off of 600~eV, a Monkhorst–Pack sampling of the Brillouin zone with a $3 \times 2 \times 3$ grid, the PBE exchange–correlation functional~\cite{Perdew1996}, and CASTEP auto-generated ultrasoft pseudo-potentials. This produced forces accurate well within an error of 0.1~eV\AA$^{-1}$, which was used as the limit tolerance for geometry optimization. Geometry optimization on the structure was performed with a LBFGS algorithm, fixing the unit cell parameters to their experimental values.

Following the UEP method, further CASTEP simulations were performed on the primitive cell of Ce$_2$Te$_5$ to obtain candidate muon sites. The presence of the muon broke the symmetry of the supercell, and convergence was more difficult. A Gaussian smearing scheme was added for the treatment of the Fermi surface, plus a Hubbard $U = 5.0$~eV for the $f$ orbital of Ce. The most probable muon site was then further refined by adopting a $2 \times 1 \times 2$ supercell and relaxing the atomic coordinates until the forces were lower than 0.1~eV/\AA$^{-1}$.
The internal field at the muon site was computed by combining the theoretical muon stopping site with the experimentally determined ferromagnetic ground state, following the approach explained in Ref.~\cite{Bonfa2018}.

NPD measurements on powdered single crystals of Ce$_2$Te$_5$ were performed at the Institut Laue–Langevin (ILL, France), using the high-intensity two-axis diffractometer D20 (high-flux mode, $\lambda = 2.42$~\AA) and the high-resolution diffractometer D2B ($\lambda = 1.594$~\AA). Low-temperature diffraction data were first collected using a dilution refrigerator, with the sample mounted in a Cu-container allowing the use of He-exchange gas to ensure proper thermalization. Measurements were carried out at several temperatures between 130~mK and 5.6~K, with data sets acquired for about 6 hours each. Due to scattering from the dilution cryostat, the copper container, sample impurities, and significant preferred orientation effects, these data are less suitable for reliable Rietveld refinement and are used only to track the temperature evolution of the magnetic scattering. The sample was subsequently transferred to a standard vanadium container and measured in a conventional cryostat for 9 hours at 1.7~K and 8~K in order to maximize the magnetic signal in the difference pattern.

Additional high-resolution diffraction measurements were performed on the D2B diffractometer to assess preferred orientation effects. Further, neutron diffraction measurements were carried out using a dilution fridge down to 0.05~K on the time-of-flight WISH neutron diffractometer at ISIS Neutron and Muon Source,~\cite{Chapon29042011} using the same powdered sample of Ce$_2$Te$_5$ mounted in the same Cu-can under He-exchange gas.

INS experiments on crushed crystals of Ce$_2$Te$_5$ were performed on the MARI time-of-flight spectrometer at the ISIS Neutron and Muon Source, UK. Approximately 2.6~g of Ce$_2$Te$_5$ powder was used for the measurements. The sample was mounted in an aluminium sample can and cooled using a closed-cycle refrigerator to access low temperatures. INS data were collected using multiple incident neutron energies to probe the CEF excitations over a wide energy range. Measurements were carried out with incident energies of $E_i$ = 8, 40, and 60~meV, using a Gd Fermi chopper with frequencies 150, 400, 400~Hz respectively. For $E_i$ = 40 and 60~meV, we used repetition rate multiplication (RRM) mode of MARI, which also gave the data of $E_i$ = 14.09, 7.08~meV (in 40~meV) and $E_i$ = 17.67, 8.32~meV (in 60~meV) in the same measurements. The majority of the low-temperature measurements were performed at $T = 8$~K (above the magnetic ordering), ensuring that the Ce$^{3+}$ ions predominantly populate the CEF ground-state Kramers doublet. Additional measurements at $T = 100$~K and 200~K were carried out with $E_i = 40$~meV to examine the temperature dependence of the inelastic response.

\section {Experimental and simulation results and discussion}

\subsection{Muon spin relaxation}
\begin{figure*}
		\includegraphics[width=15 cm, keepaspectratio]{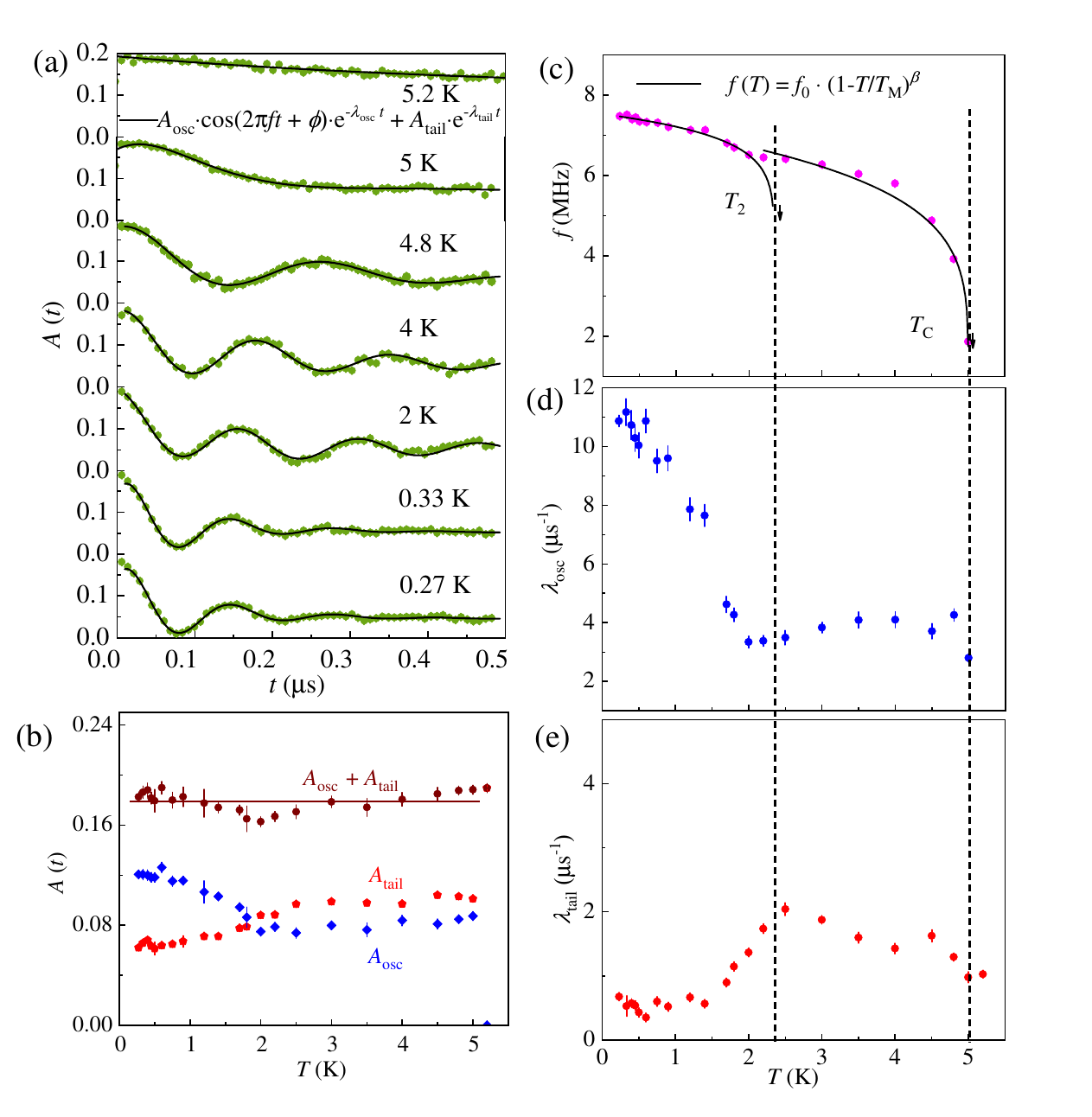}
		\caption{(a) ZF-$\mu$SR spectra for Ce$_2$Te$_5$ at different temperatures. Solid lines show fitting to Eq.~(\ref{eq1}). (b) Asymmetries as a function of temperature and their sum. (c) Temperature dependence of the muon precession frequency associated with the static internal field at the muon site. The line shows fitting to a power law. (d)-(e) Temperature dependence of the relaxation rates $\lambda_{\rm osc}$ and $\lambda_{\rm tail}$. The vertical dashed lines are guides to the eye.}
		\label{ZFMUSR}
\end{figure*}

To investigate the nature of the ground state of Ce$_2$Te$_5$ at a microscopic scale and compare it with the low-temperature thermal- and transport measurements, we conducted ZF-$\mu$SR measurements down to 0.27~K. $\mu$SR is an effective local-probe technique, capable of detecting small magnetic moments, down to 0.005~$\mu_{\text{B}}$, and to distinguish between random static fields associated with phenomena like muon dipolar coupling and quasi-static nuclear moments, as well as fluctuating magnetic fields arising from electronic-spin fluctuations. For more information about the $\mu$SR technique, refer to Refs.~\cite{MUSR,2022msai.bookB}.
To investigate the muon spin rotation and depolarization changes, we collected time-dependent ZF-$\mu$SR spectra of Ce$_2$Te$_5$ at various temperatures ranging from 0.27\,K to 5.2\,K. Representative ZF time spectra at different temperatures in a short time frame are displayed in Fig.~\ref{ZFMUSR}(a). The spectrum at 5.2\,K exhibits only slow relaxation with no oscillations, indicating the absence of a magnetic order, consistent with the paramagnetic state expected above the magnetic transition. In contrast, clear oscillations of the muon-spin polarization can be seen at temperatures below 5.2\,K, indicating the Larmor precession of muon spins in the static internal field, typical of magnetic ordering. The angular frequency $\omega$ of these oscillations is dependent on the local magnetic field $B$ at the muon stopping site, through the relation $\omega=\gamma_\mu B$. Here, the gyromagnetic ratio of muons is given by $\gamma_\mu/2\pi=135.5$~MHz/T. 

The ZF-$\mu$SR spectra could be successfully fitted by an exponentially relaxing oscillating function: 
\begin{equation}
\begin{split}
A_{\mathrm{ZF}}(t)=A_{\mathrm{osc}} \cos \left(2 \pi f t+\phi\right) \cdot e^{\left(-\lambda_{\mathrm{osc}} t\right)}\\
+ A_{\text {tail }} \cdot e^{\left(-\lambda_{\text {tail}} t\right)},
\end{split}
\label{eq1}
\end{equation}
\noindent where the first term represents the oscillation of the muon spin polarization (called transverse component) coming from the static local field. $A_{\text{osc}}$ is the amplitude of the precessing component, $2\pi f$ is the angular frequency of the Larmor precession, associated with the local field, while $\phi$ is a phase offset and $\lambda_{\text{osc}}$ is the relaxation rate of the cosine function. The second term is the non-oscillating component associated with the muons experiencing local fields parallel to the initial muon spin direction, or strongly damped fields due to a distribution of internal fields and/or slow spin fluctuations. $A_{\text{tail}}$ and $\lambda_{\text{tail}}$ are the asymmetry and relaxation rate of this non-oscillating component (called longitudinal component). 

The fits to the spectra by Eq.~(\ref{eq1}) are shown by the solid curves in Fig.~\ref{ZFMUSR}(a). The temperature dependencies of the fitting parameters determined from the best fits are illustrated in Figs.~\ref{ZFMUSR}(b)-\ref{ZFMUSR}(e).

Figure~\ref{ZFMUSR}(b) shows the temperature dependence of the initial asymmetries $A_{\text{osc}}$ and $A_{\text{tail}}$. While the total asymmetry $A_{\text{osc}}$ + $A_{\text{tail}}$ remains nearly constant over the entire temperature range, a pronounced redistribution of asymmetry between the two components is observed upon cooling. Specifically, $A_{\text{osc}}$ increases whereas $A_{\text{tail}}$ decreases with decreasing temperature. This behavior indicates that the full sample volume remains magnetic, but the local magnetic field environment at the muon stopping sites evolves with temperature. Such behavior can arise from multiple muon stopping sites, long-range bulk ordering, short-range or spatially inhomogeneous magnetic correlations, or the progressive slowing down of spin fluctuations upon cooling.

Analysis of the temperature-dependent oscillation frequency provides insight into the spin dynamics of the system. The extracted precession frequency from the refinement as a function of temperature is shown in Fig.~\ref{ZFMUSR}(c), revealing a gradual increase in the precession frequency as the temperature is lowered, suggesting a second-order-like behavior of the magnetic phase transition.

The most striking feature of these results is that the muon precession frequency exhibits subtle changes in slope near the characteristic temperatures reported from bulk measurements. To parameterize these changes, the temperature dependence of the frequency was analyzed using the phenomenological power-law form
\begin{equation}
f(T) = f_0\,(1-T/T_{\mathrm{M}})^\beta ,
\label{Eq2}
\end{equation}
where $T_{\mathrm{M}}$ represents the characteristic temperature associated with a given temperature interval. As shown in Fig.~\ref{ZFMUSR}(c), the frequency shows a change in its temperature dependence around $T_{\mathrm{2}}$. Independent fits to these intervals yield characteristic temperatures of $T_{\mathrm{C}} = 5.0(1)$~K and $T_{\mathrm{2}} = 2.3(2)$~K, with the corresponding fitting parameters summarized in Table~\ref{table1}. The highest-temperature transition at $T_{\mathrm{C}}$ corresponds to the onset of long-range ferromagnetic order, consistent with bulk magnetization and neutron diffraction results, while the lower temperature $T_{\mathrm{2}}$ marks an additional anomaly within the magnetically ordered state. Because of the limited number of data points and the relatively small variation of the precession frequency at the lowest temperatures, it is difficult to independently track the low-temperature behavior below $\sim1$ K, with sufficient statistical confidence. We therefore fitted the data from the lowest measured temperature up to $T_{\mathrm{2}}$ using a common phenomenological form to characterize the overall evolution of the internal field in this temperature range, rather than assigning an additional transition at $T_{\mathrm{3}}$.

The critical exponent extracted near $T_{\mathrm{C}}$, $\beta = 0.19(1)$, is substantially smaller than the values expected for conventional three-dimensional universality classes ($\beta \approx 0.33$). Upon further cooling, the effective exponent decreases to even smaller effective values of $\beta \approx 0.08$ in the temperature ranges associated with $T_{\mathrm{2}}$. These small effective exponents are not consistent with the true asymptotic critical behavior of any established universality class, but rather reflect the local response of muons to a distribution of internal fields in the ordered state. Because $\mu$SR measures the local internal field at the muon sites rather than the macroscopic order parameter, such effective scaling exponents can differ significantly from the bulk critical exponents~\cite{PDalmas_1997}. In fact, bulk magnetization measurements on Ce$_2$Te$_5$ yield $\beta\approx0.31$, consistent with three-dimensional magnetic criticality, whereas the $\mu$SR-derived effective exponent reflects the crossover behavior within the already ordered state.

Furthermore, in the $B_{\rm int}(T)$ representation, the parameter $B_0$ corresponds to the $T\!\rightarrow\!0$ extrapolated internal field at the muon site, i.e., the saturation field associated with the static ordered state. Using $B_{\rm int}=f/(\gamma_\mu/2\pi)$, the fitted $f_0\simeq 7.3$--7.5~MHz yields $B_0 \simeq 54$--55~mT. The fact that $B_0$ remains nearly unchanged across the different temperature intervals indicates that the low-temperature state reaches essentially the same saturation internal field, while the anomalies at $T_{\rm 2}$ primarily modify the temperature evolution (effective critical behavior and/or field distribution) rather than introducing a distinct long-range ordered phase. 

The temperature dependencies of the relaxation rates for the oscillating and non-oscillating components $\lambda_{\text{osc}}$ and $\lambda_{\text{tail}}$ are shown in Figs.~\ref{ZFMUSR}(d) and \ref{ZFMUSR}(e), respectively. Both $\lambda_{\text{osc}}$ and $\lambda_{\text{tail}}$ show a cusp-like enhancement around the magnetic transition region, reflecting the spin dynamics of the localized Ce moments in the system and resulting in a broadening of the field distribution width at the muon sites. This behavior is characteristic of critical spin fluctuations in the vicinity of a magnetic phase transition.

Moreover, combining $B_0\approx 55$~mT with the ordered moment $\mu_{\rm ord}\approx 0.8~\mu_\mathrm{B}$ from neutron diffraction implies an effective coupling $A_\mu^{\rm eff}\approx 0.07$~T/$\mu_\mathrm{B}$, consistent with a dipolar field scale for a muon stopping site a few \AA\ from the ordered Ce sublattice~\cite{PDalmas_1997,Yaouanc2011}. The muon stopping site predicted by the simulations is indicated with a blue sphere in Fig.~\ref{fig:MuStopSite} in the appendix. The calculated internal field at the muon stopping site using the dipolar field contribution and the magnetic structure from neutron diffraction (see Sec.~B) is $B_{\rm int} = 53$~mT, which corresponds to a frequency of $f_0\simeq 7.2$~MHz and is in excellent agreement with the experimental results. The distance between the muon and its four nearest Ce neighbors is 3.36~\AA.

\begin{table}
	\centering
	\caption {Parameters obtained from fitting the temperature dependence of the muon precession frequency using Eq.~(\ref{Eq2}).}
	\vskip 0.5cm
	\addtolength{\tabcolsep}{+10.0pt}
	\begin{tabular}{c c c c}
	\hline
   & $f_0$ & $T_{\text{M}}$ & $\beta$ \\[1ex]
$2.0\leq T\leq5.2$~K & 7.3(1) & 5.0(1) &  0.19(1)   \\[1ex]
$T <2.0$~K & 7.52(2) & 2.3(1) & 0.08(1) 	\\[1ex]	
            \hline
		\hline
	\end{tabular}
	\label{table1}
\end{table}

\subsection{Neutron diffraction}

 \begin{figure*}

		\includegraphics[width=15cm]{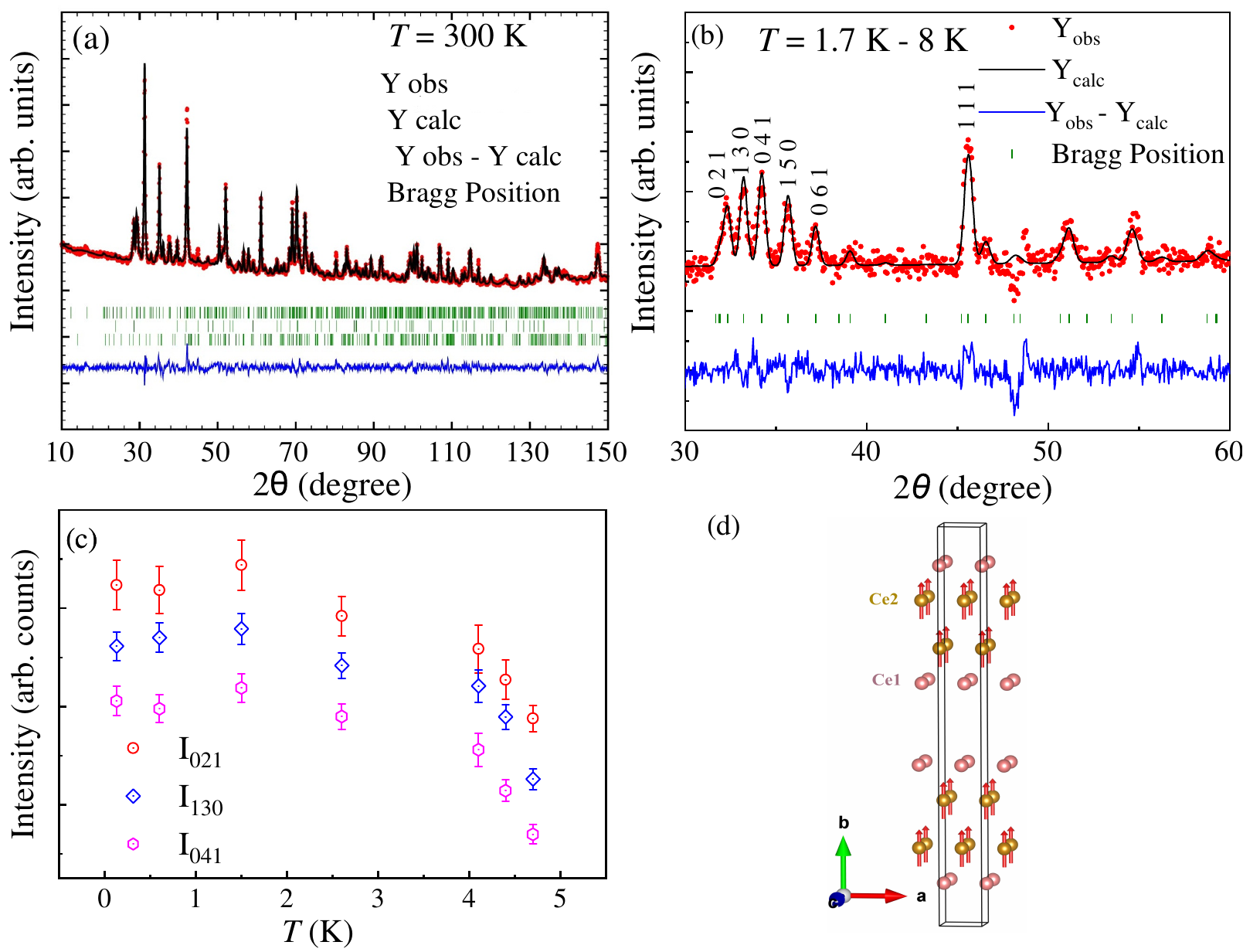}
		\caption{Refinement of the diffraction pattern of Ce$_2$Te$_5$ (a) at 300~K in the paramagnetic state and (b) at the temperature-difference data 1.7 – 8~K using the fixed scale factor from the refinement of the paramagnetic data at 8~K. The red symbols show the experimental data. The solid black line shows the refined fit, vertical green ticks show the position of the Bragg reflections and the bottom blue line shows the difference between the observed and calculated diffraction patterns. (c) Temperature dependence of the integrated intensities of the magnetic Bragg reflections (021), (130), and (041), illustrating the development of the magnetic order parameter below $T_{\mathrm C}$. (d) Magnetic structure of Ce$_2$Te$_5$ obtained from neutron powder diffraction refinement of the difference pattern ($T = 1.7$~K -- $8$~K). The ordered magnetic moments are aligned ferromagnetically along the crystallographic $b$ axis and reside only on the Ce2 site, while the Ce1 site carries a negligible magnetic moment within experimental uncertainty. For clarity, the magnetic structure is shown using an extended unit cell along the $a$ and $c$ directions. The crystallographic axes are indicated.}
		\label{ND}
	
\end{figure*}

To follow the development of magnetic order across the entire temperature range, NPD data were collected from room temperature down to 0.05~K. A comparison of data collected below and above the ordering temperature reveals additional scattering of magnetic origin at low temperatures (see Fig.~\ref{NDS1} in the appendix), whereas the high temperature pattern contains only the nuclear contribution. Importantly, the positions of the magnetic reflections do not change with temperature, indicating a single magnetic propagation vector over the entire ordered regime (Fig.~\ref{NDS1} in the appendix). The magnetic intensity remains essentially unchanged between 130~mK and $\sim$2.6~K, and starts to decrease only above $\sim$4.1~K. At 5.1~K, the difference pattern reveals only very weak, broad residual magnetic scattering, while at 5.6~K no magnetic signal remains within the sensitivity of the present experiment. 

Figure~\ref{ND}(a) shows the D20 pattern collected at 300~K, i.e., in the paramagnetic state, used to refine the nuclear structure and determine the scale factor. The refinement confirms that Ce$_2$Te$_5$ adopts an orthorhombic crystal structure belonging to the space group $Cmcm$ (No.~63). The refined lattice and atomic parameters are summarized in Tables~\ref{TS1} and \ref{TS2}. Figure~\ref{ND}(b) and Fig.~\ref{NDS2} in the Appendix  display the difference patterns, measured on the D20 (1.7 - 8~K) and WISH (0.05 - 10~K) instruments, respectively, between the low- (ordered) and high- (disordered) temperature phases, highlighting the magnetic contribution. In  both cases, the magnetic reflections could be indexed with a commensurate propagation vector $\mathbf{k}=(0,0,0)$ and their intensities were modelled using the \textsc{FullProf} program~\cite{RodriguezFullProf,Rodriguez2025}. Magnetic symmetry analysis was carried out with \textsc{BasIreps}~\cite{RodriguezFullProf,Rodriguez2025,ritter2011}, and the subsequent refinement of the magnetic difference pattern yielded a ferromagnetic arrangement with moments aligned along the crystallographic $b$ axis (Fig.~\ref{ND}(d)). The magnetic ordering implies the $Cm'cm'$ magnetic space group, which keeps the lattice vectors and origin of the paramagnetic structure. Remarkably, only one of the two crystallographically inequivalent Ce sites, namely Ce2, carries a substantial ordered moment, $m \approx 0.85(3)~\mu_{\mathrm{B}}$ per Ce at 0.05~K, while the moment on the Ce1 site is negligible, within the refinement sensitivity. 

To quantify the temperature evolution of the magnetic order, the integrated intensities of three representative magnetic reflections, namely (021), (130), and (041), were extracted and are plotted in Fig.~\ref{ND}(c). All three reflections exhibit a similar temperature dependence, remaining nearly constant at low temperatures before decreasing on approaching the magnetic transition. The magnetic intensity decreases continuously on warming and is strongly suppressed near 5~K, consistent with the ferromagnetic ordering temperature $T_{\mathrm C}\approx 5$~K. Importantly, no additional anomaly is observed in the temperature dependence of the magnetic Bragg intensities at $T_2$, indicating that no detectable change occurs in the long-range magnetic order parameter at $T_2$ within the sensitivity of the present experiment.

This observation is particularly noteworthy because pronounced anomalies have been reported at $T_2$ and lower temperatures in the electrical resistivity $\rho(T)$ and specific heat $C_p(T)$~\cite{PhysRevMaterials.6.094407}, suggesting additional changes in the low-energy electronic and magnetic properties of Ce$_2$Te$_5$. However, NPD places important constraints on the nature of these anomalies. Importantly, neither the positions nor the relative intensities of the magnetic Bragg reflections show any observable variations throughout the ordered temperature regime. The data therefore provide no evidence for a modification of the magnetic propagation vector or for a significant change of the long-range magnetic structure below $T_{\mathrm{C}}$. The strong sublattice selectivity of the magnetic ordering, characterized by an ordered moment on the Ce2 site and no detectable ordered moment on the Ce1 site, is an important feature of Ce$_2$Te$_5$. However, the microscopic origin of this sublattice selectivity, as well as the precise nature of the additional transitions observed below 5 K in the heat capacity ~\cite{Chen2017, PhysRevMaterials.6.094407}, cannot be established from the present measurements.

\subsection{Inelastic neutron scattering}

\begin{figure*}
	\includegraphics[width=15cm, keepaspectratio]{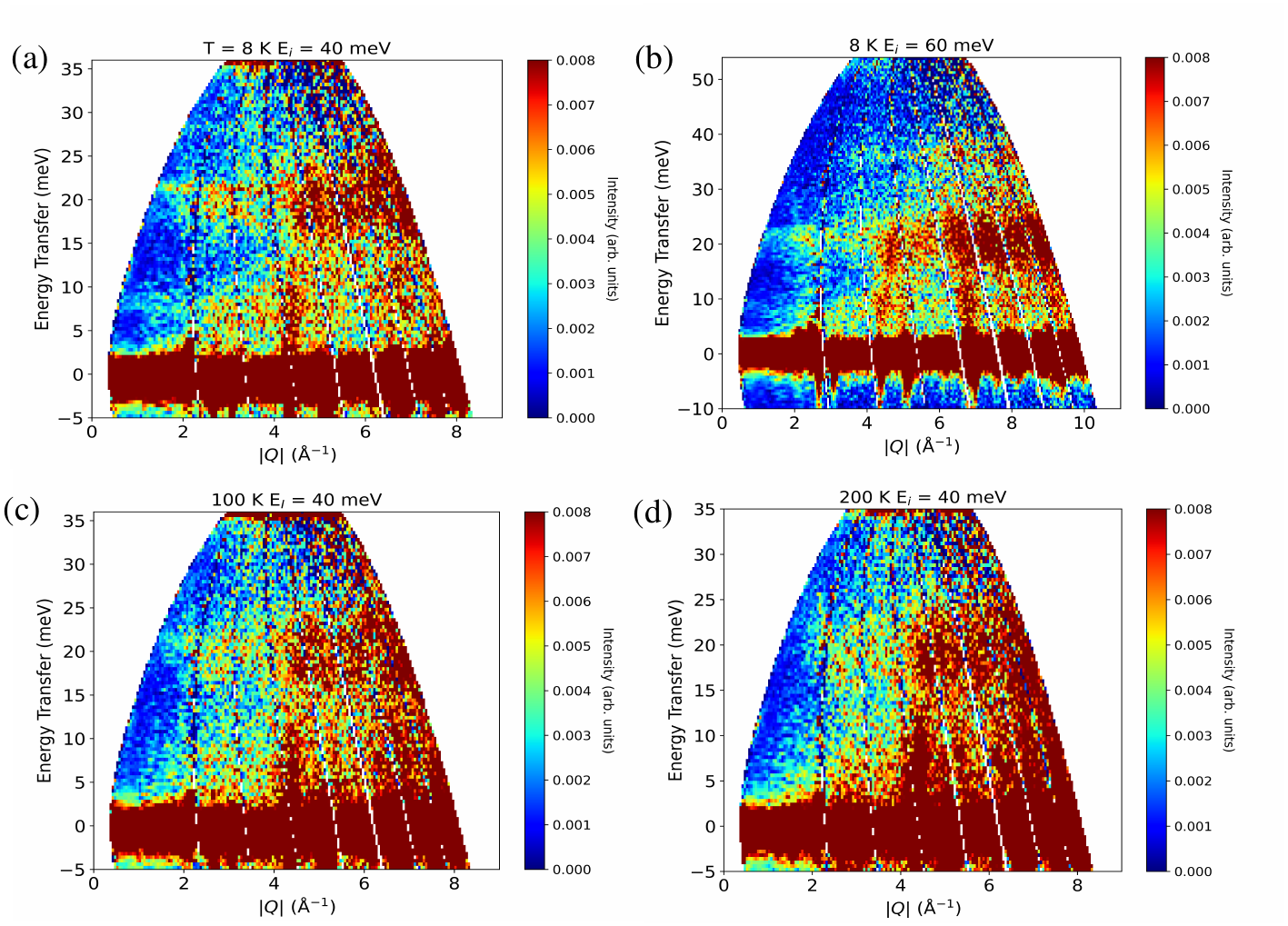}
	\caption{Color-coded INS intensity maps of crushed Ce$_2$Te$_5$ crystals, plotted as a function of energy transfer and momentum transfer $|Q|$. Panels show data collected at (a) $T = 8$~K with $E_i = 40$~meV, (b) $T = 8$~K with $E_i = 60$~meV, (c) $T = 100$~K with $E_i = 40$~meV, and (d) $T = 200$~K with $E_i = 40$~meV. The intensity is shown in arbitrary units.}
	\label{INS1}
\end{figure*}

\begin{figure}
\includegraphics[width=8.0 cm, keepaspectratio]{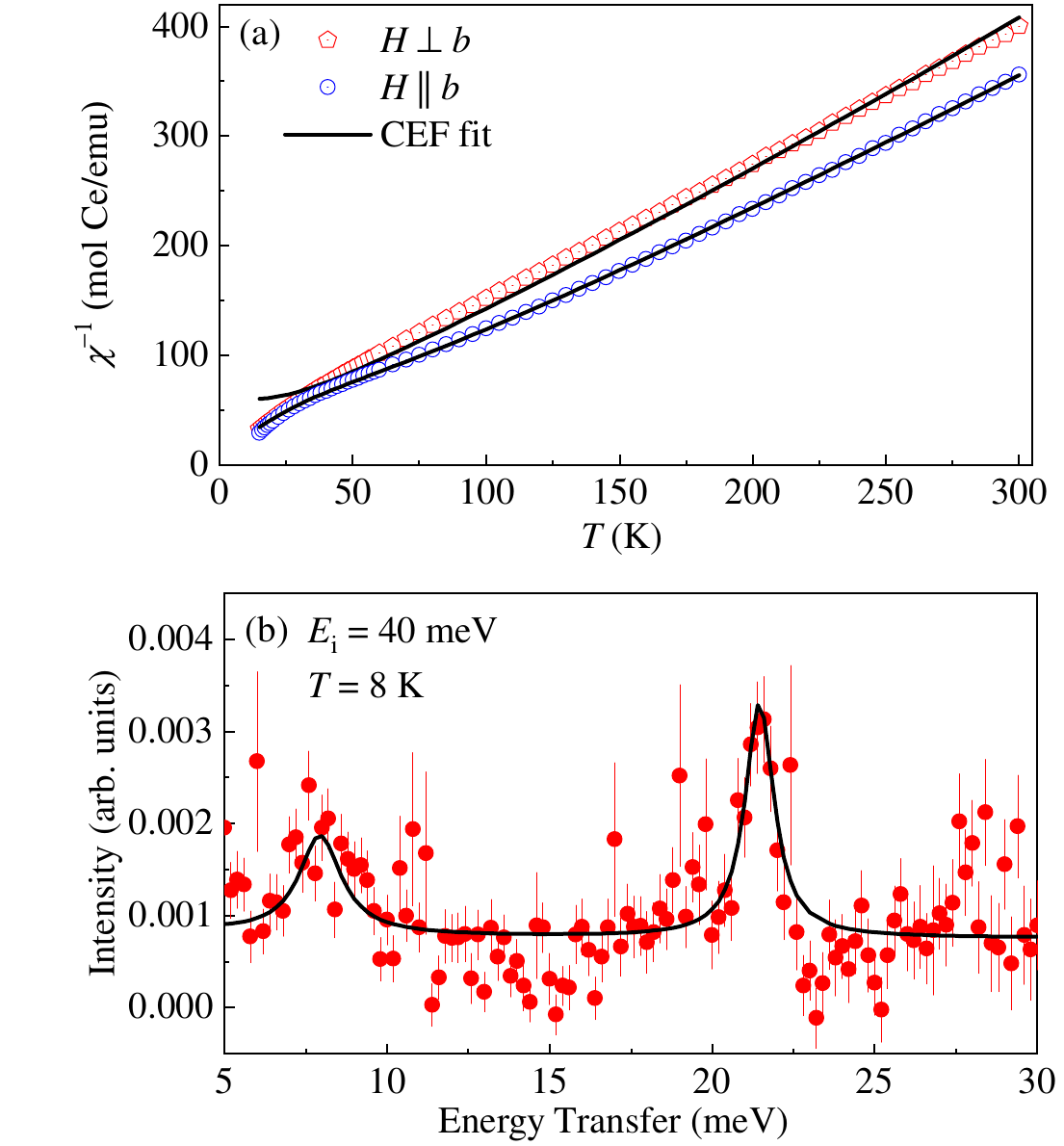}
\caption{
(a) Inverse magnetic susceptibility $\chi^{-1}(T)$ of Ce$_2$Te$_5$ single crystal measured for magnetic field applied parallel ($H \parallel b$) and perpendicular ($H \perp b$) to the crystallographic $b$ axis. Solid lines represent fits based on the CEF model (see text). (b) Magnetic INS spectrum of Ce$_2$Te$_5$ at $T = 8$~K measured with an incident neutron energy $E_i = 40$~meV, integrated over low momentum transfer, $Q=0-3$~\AA$^{-1}$. Solid lines show the CEF model calculation.}
\label{INS2}
\end{figure}

INS measurements were performed to investigate the CEF excitations of the Ce$^{3+}$ ($4f^1$) ions in Ce$_2$Te$_5$. Figures~\ref{INS1}(a)-(d) show color-coded INS intensity maps of Ce$_2$Te$_5$ as a function of energy transfer and momentum transfer $|Q|$ for different incident energies and temperatures. At low temperature ($T = 8$~K), enhanced scattering intensity is observed at low $|Q|$ near 8~meV and 21~meV (see Figs.~\ref{INS1}(a)-(b)), consistent with a magnetic origin governed by the Ce$^{3+}$ magnetic form factor. At higher temperatures (100 and 200~K), the magnetic signal becomes increasingly masked by phonon scattering and statistical noise, reflecting the weak magnetic cross section (see Figs.~\ref{INS1}(c)-(d) and Appendix Figs.~\ref{INS3}(c)-(d)).

Figure~\ref{INS2}(b) shows the INS intensity as a function of energy transfer after integrating over a low-$Q$ range to enhance the magnetic contribution. Due to the absence of a nonmagnetic reference compound (such as La$_2$Te$_5$), the phonon background was estimated using a standard low-$Q$ minus scaled high-$Q$ subtraction procedure. While this approach reduces the dominant phonon scattering, a residual background remains, particularly at low energies. At $T = 8$~K, the INS spectrum reveals a clear magnetic excitation centered near $\sim 8$ and 21~meV. The magnetic origin of these excitations is supported by their enhanced intensity at low momentum transfer, consistent with the Ce$^{3+}$ magnetic form factor as well as the decreases in the intensity at 100~K and 200~K (see appendix Figs.~\ref{INS3}(c)-(d)). 

To analyze the INS data quantitatively, the CEF Hamiltonian for Ce$^{3+}$ ($J = 5/2$) was modeled using the Stevens operator formalism. Although Ce$_2$Te$_5$ crystallizes in the orthorhombic $Cmcm$ structure, the accidental near-degeneracy of the crystallographic $a$ and $c$ directions gives rise to an approximately tetragonal local environment around the Ce ions. Consistent with the magnetic anisotropy reported previously~\cite{PhysRevMaterials.6.094407}, the local point symmetry was therefore approximated by $C_{4v}$, with the fourfold rotational axis oriented along the crystallographic $b$ axis. Under this approximation, the CEF Hamiltonian is described by three independent Stevens parameters, $B_2^0$, $B_4^0$, and $B_4^4$. 

Furthermore, despite the presence of two crystallographically inequivalent Ce sites (Ce1 and Ce2), the INS spectra reveal only two well-defined CEF excitations rather than the four excitations expected for two distinct CEF schemes. We therefore adopted a single-site CEF model, assuming that the observed magnetic response is dominated by one Ce environment. This assumption is also consistent with the NPD results, which indicate that only the Ce2 site carries a sizeable ordered magnetic moment. The corresponding CEF Hamiltonian is given by

\begin{equation}
H_{\mathrm{CEF}} = B_{2}^{0} O_{2}^{0} + B_{4}^{0} O_{4}^{0} + B_{4}^{4} O_{4}^{4},
\end{equation}

where $B_n^m$ are the CEF parameters and $O_n^m$ are the Stevens operators~\cite{Stevens_1952}. Within this symmetry, the sixfold degenerate $J = 5/2$ multiplet of Ce$^{3+}$ splits into three Kramers doublets, giving rise to two allowed CEF transitions from the ground state at low temperature. The $b$-axis (parallel to the $z$-axis) was taken as the quantization axis. The CEF parameters were obtained by simultaneously fitting the magnetic INS data and single-crystal magnetic susceptibility data~\cite{PhysRevMaterials.6.094407} using the Mantid software suite~\cite{ARNOLD2014156}, with a Lorentzian lineshape applied to the inelastic excitations.

As shown in Figs.~\ref{INS2}(a) and \ref{INS2}(b), the calculated CEF model (solid lines) provides a consistent description of both datasets. In particular, the INS spectrum at $T = 8$~K is well reproduced, capturing two magnetic excitations centered at approximately 8 and 21~meV, corresponding to the two excited CEF doublets of Ce$^{3+}$. The best-fit CEF parameters (in meV) are: $B_2^0$ = --0.5497(2), $B_4^0$ = 0.0605(8), and $B_4^4$ = --0.0100(6). These values yield excited CEF doublets at $\Delta_1$ = 7.94~meV and $\Delta_2$ = 21.46~meV. The fitted molecular field constants $\lambda$ (in mol/emu) and temperature-independent susceptibilities $\chi_0$ (in emu/mol) for fields along and perpendicular to the $b$ directions are: $\lambda$ = --10(1) and 24(2); $\chi_0$ = --4.4(6) $\times$ $10^{-5}$ and --1.8(1) $\times$ $10^{-4}$, respectively. 

Using the refined CEF parameters, we calculated the eigenfunctions of the three Kramers doublets, shown below: 

\begin{equation}
\begin{split}
\Psi_0 & =
0.034\,|\pm \tfrac{5}{2}\rangle
+ 0.999\,|\mp \tfrac{3}{2}\rangle, \\
\Psi_1 & =
0.999\,|\pm \tfrac{5}{2}\rangle
- 0.034\,|\mp \tfrac{3}{2}\rangle,\\
\Psi_2 & =
|\pm \tfrac{1}{2}\rangle
\end{split}
\end{equation}

The ground-state Kramers doublet is predominantly of $|\mp 3/2\rangle$ character, with only a small admixture of the $|\pm 5/2\rangle$ state. The first excited doublet, located at 7.94~meV, is primarily of $|\pm 5/2\rangle$ character, while the second excited doublet at 21.46~meV corresponds to the pure $|\pm 1/2\rangle$ state. The resulting CEF scheme accounts for the pronounced single-ion anisotropy of Ce$^{3+}$ ions observed in the high-temperature paramagnetic regime.

In addition to the two robust magnetic excitations at approximately 8~meV and 21~meV, the INS spectra show a weak and broad intensity enhancement at higher energies around $\sim 27$~meV in the 40 and 60~meV data sets. However, this feature lies in an energy range where the magnetic INS signal is strongly suppressed and the spectra are dominated by phonon background, which cannot be fully removed in the absence of a nonmagnetic reference compound. 
Moreover, its intensity does not exhibit a clear and systematic temperature or momentum-transfer dependence characteristic of a well-defined CEF excitation. We therefore refrain from assigning this weak feature to an additional CEF transition. Within the experimental resolution and signal-to-noise ratio of the present measurements, the INS data provide firm evidence only for two CEF excitations below $\sim 25$~meV, consistent with the splitting of the Ce$^{3+}$ $J=5/2$ multiplet into three Kramers doublets under $C_{4v}$ symmetry.

\subsection{Discussion}

NPD provides the most direct constraint on the static long-range order in Ce$_2$Te$_5$. The difference patterns (low $T$ - 8~K) are indexed with a commensurate propagation vector $\mathbf{k}=(0,0,0)$ and are consistently described, over the full ordered regime accessed here by a ferromagnetic arrangement of moments along the crystallographic $b$ axis. The refinement further resolves a pronounced sublattice selectivity: only one of the two crystallographically inequivalent Ce sites (Ce2) carries an ordered moment ($m\simeq0.85(3)~\mu_{\mathrm B}$/Ce at $\sim$1.7~K), whereas the second Ce site (Ce1) remains moment-free within the experimental sensitivity. Importantly, upon cooling from $\sim$5~K to the base temperature (0.13~K on D20 and 0.05~K on WISH), no additional magnetic Bragg reflections, peak splittings, or peak-position shifts are observed, demonstrating that the magnetic structure retains the same propagation vector and symmetry throughout the ordered state. This directly supports the bulk assignment of the primary transition as a ferromagnetic instability with easy-$b$ anisotropy and provides a microscopic realization of the “one-sublattice-dominant” ordering scenario discussed previously~\cite{PhysRevMaterials.6.094407}.

Zero-field $\mu$SR, in contrast, resolves an additional anomaly below $T_{\mathrm C}$ through a change in the local internal field scale and relaxation channels at $T_{\mathrm{2}}$. Converting the fitted saturation frequencies ($f_0\simeq7.3$ to $7.5$~MHz) to internal fields using $B_0 = (2\pi f_0)/\gamma_\mu$ yields $B_0 \approx 54-55$~mT. The near-constancy of $B_0$ across the different temperature intervals indicates that the dominant ordered-moment amplitude sensed at the muon site is essentially unchanged once long-range order sets in, consistent with the temperature-independent magnetic Bragg intensities deep in the ordered state. Therefore, the lower-temperature feature at $T_{\mathrm{2}}$ is most naturally attributed not to the emergence of a new propagation vector or of a distinct long-range magnetic phase, but rather to more subtle modifications of the local magnetic environment that remain below the sensitivity of the present neutron diffraction experiment. Such effects may include changes in the local field distribution, magnetic fluctuations, short-range correlations, or electronic degrees of freedom that influence the muon relaxation channels without producing detectable changes in either the magnetic Bragg peak positions or their relative intensities. In this context, the pronounced anomalies observed in $\rho(T)$ and $C_p(T)$ at $T_{\mathrm{2}}$ and below remain compatible with a single magnetic structure determined by diffraction, while highlighting the presence of additional low-energy processes that are more readily detected by local and thermodynamic probes.

A key point for comparison with the critical analysis in the bulk study is that the exponent $\beta$ extracted from $\mu$SR internal-field fits is not, in general, identical to the thermodynamic order-parameter exponent obtained from $M(H,T)$ scaling. In Ce$_2$Te$_5$, the bulk critical behavior around $T_{\mathrm C}$ was found to be three-dimensional with $\beta\simeq0.31$ and $\gamma\simeq0.99$, close to the 3D Ising-like $\beta$, while $\gamma$ remains reduced, likely reflecting long-range RKKY interactions and strong anisotropy.
By contrast, $\mu$SR measures a local field $B_\mu(T)$ that is a weighted dipolar/contact sum over the ordered moments and can be affected by (i) nonuniform order parameters on inequivalent Ce sublattices, (ii) temperature-dependent canting angles and domain populations, or (iii) additional relaxation channels from slow fluctuations. These effects can renormalize the apparent power-law exponent in $B_\mu(T)\propto(1-T/T_{\mathrm M})^\beta$ even when the underlying long-wavelength universality class is 3D. Consequently, the reduced $\mu$SR-derived $\beta$ values at lower temperatures are best interpreted as signatures of crossover/rearrangement behavior within the ordered state (consistent with NPD), rather than as evidence for a distinct universality class competing with the 3D criticality established at $T_{\mathrm {C}}$ by bulk scaling. 

Finally, the INS results place the relevant single-ion CEF energy scale in the $\sim$8--22~meV window and reproduce the strong easy-$b$ anisotropy expected for the Ce$^{3+}$ ground state. The refined CEF scheme yields a ground-state Kramers doublet that is overwhelmingly dominated by the $|\mp 3/2\rangle$ component, resulting in a highly anisotropic single-ion magnetic response. With the CEF quantization axis oriented along the crystallographic $b$ direction, the magnetic moment is maximized along $b$, while the responses along the $a$ and $c$ directions are strongly suppressed. This Ising-like anisotropy is consistent with the pronounced susceptibility anisotropy reported below $T_{\mathrm C}$ and naturally explains the moment direction determined by neutron diffraction. The ferromagnetic arrangement with propagation vector $\mathbf{k}=(0,0,0)$ then arises from the exchange interactions acting between these anisotropic Ce moments.

\section{Conclusions}

In summary, we have investigated the magnetic ground state of the layered vdW $f$-electron system Ce$_2$Te$_5$ using $\mu$SR, NPD, and INS. Zero-field $\mu$SR measurements reveal the onset of static magnetism below $T_{\mathrm{C}} = 5.0(1)$~K, accompanied by an additional low-temperature anomaly at $T_{\mathrm{2}} = 2.3(2)$~K, consistent with features reported previously in bulk thermodynamic and transport measurements. Analysis of the temperature dependence of the internal field yields reduced critical exponents, reflecting the influence of reduced dimensionality, strong anisotropy, and competing magnetic interactions. In contrast, NPD data collected between 0.05~K and 8~K reveal a single long-range ordered magnetic phase with no detectable changes in the magnetic Bragg peak positions across the entire temperature range. The magnetic structure is characterized by a commensurate propagation vector $\mathbf{k}=(0,0,0)$ with a ferromagnetic alignment of the ordered moments along the crystallographic $b$ axis. Notably, only one of the two crystallographically inequivalent Ce sites (Ce2) carries an ordered moment, with a magnitude of $0.85(3)~\mu_{\mathrm{B}}$ per Ce atom at 1.7~K. The apparent discrepancy between the multiple anomalies detected by local/bulk probes and the single magnetic phase observed by neutron diffraction indicates that the low-temperature feature at $T_{\mathrm{2}}$ does not necessarily correspond to additional symmetry-breaking transitions. Instead, they likely originate from subtle modifications of the magnetic state that remain below the sensitivity of neutron diffraction experiments. These include changes in the local magnetic field distribution, magnetic fluctuations, short-range correlations, or other low-energy electronic and magnetic degrees of freedom. INS measurements further establish the CEF energy scale and indicate a strong single-ion anisotropy of Ce$^{3+}$, providing the microscopic basis for the observed magnetic behavior. Together, these results demonstrate that Ce$_2$Te$_5$ hosts a robust ferromagnetic ground state with complex low-temperature magnetic responses arising from the interplay of layered structure, CEF effects, and competing exchange interactions. To determine the nature of the 27-meV excitation, i.e., whether it is due to a CEF excitation of the second Ce site or to phonon excitations, future single-crystal INS experiments using polarized neutrons are desirable.

\begin{acknowledgments}
We gratefully acknowledge the ISIS Neutron and Muon Source for the beam time on MARI (RB2220773, DOI: 10.5286/ISIS.E.RB2220773))  and on WISH (RB2310667, https://doi.org/10.5286/ISIS.E.RB2310677-1) and ILL for beam time on D20/D1B/D2B (Exp. No. 5-31-2931). R.T. acknowledges support from the Science and Engineering Research Board (SERB) under the National Postdoctoral Fellowship (PDF/2022/001539) and from the Indian Nano Mission through a postdoctoral fellowship. DTA would like to thank the Royal Society of London for International Exchange funding between the UK and Japan, Newton Advanced Fellowship funding between UK and China, EPSRC UK for the funding (Grant No. EP/W00562X/1) and the CAS for PiFi 2025PG0007 Fellowship. LL acknowledges financial support from the Ada Lovelace Centre, a centre of expertise in scientific software based at the Scientific Computing Department in STFC. Work at Los Alamos National Laboratory were performed under the auspices of the U.S. Department of Energy, Office of Basic Energy Sciences, Division of Materials Science and Engineering under project “Quantum Fluctuations in Narrow-Band Systems”.

\end{acknowledgments}

\appendix 

\section{Visualization of the Muon Stopping Site}
\begin{figure}[!ht]
    \centering
    \includegraphics[width=3cm]{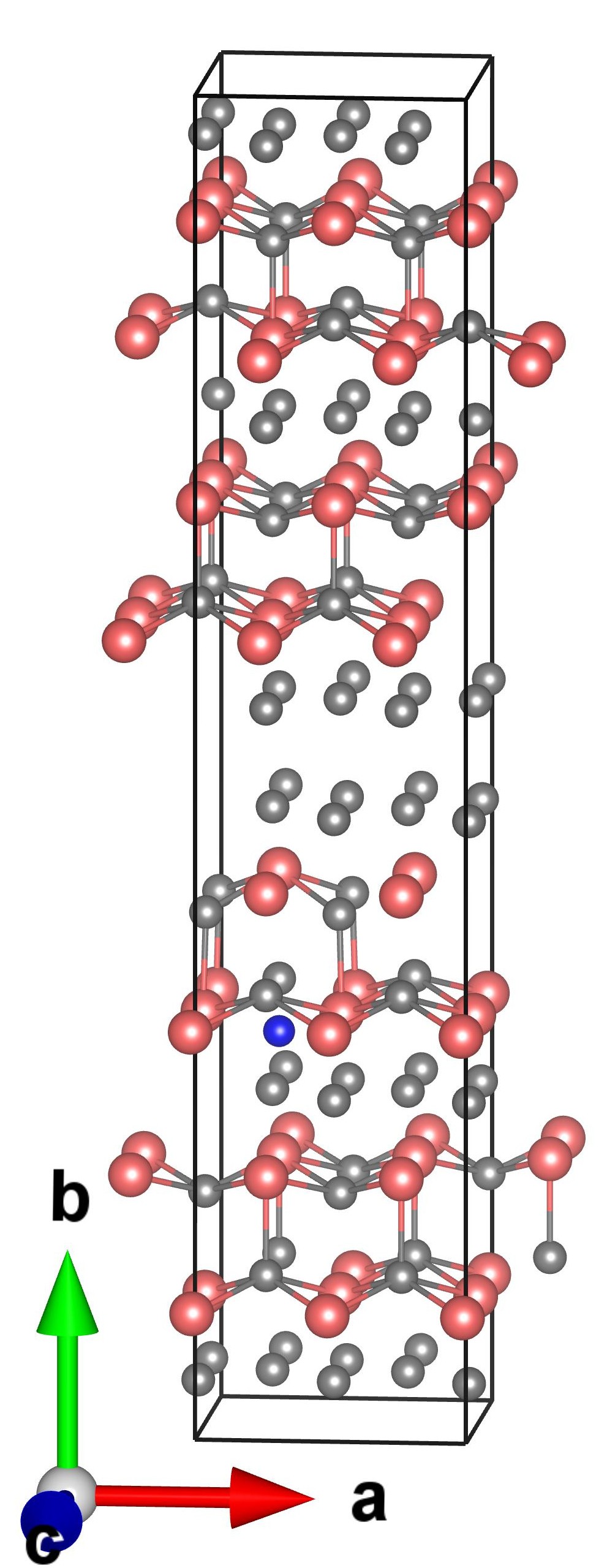}
    \caption{The calculated muon stopping site is indicated by a blue sphere in a $2 \times 1 \times 2$ supercell. The calculated internal field at the muon site is $B_{\rm int} = 53$~mT, and the distance between the muon and its four nearest Ce neighbors is 3.36~\AA.}
    \label{fig:MuStopSite}
\end{figure}

Figure~\ref{fig:MuStopSite} displays the calculated muon stopping site in Ce$_2$Te$_5$. The site was obtained from density-functional-theory calculations using a relaxed $2 \times 1 \times 2$ supercell. The resulting relaxed muon position is located at fractional coordinates (x,y,z) = (0.2498,\,0.2846,\,0.3752) in the supercell. The corresponding internal-field calculations were performed using the relaxed supercell structure to include the effect of local distortions induced by the implanted muon.

\section{Supplementary Neutron Diffraction and Structural Data}
\begin{figure}

		\includegraphics[width=9cm]{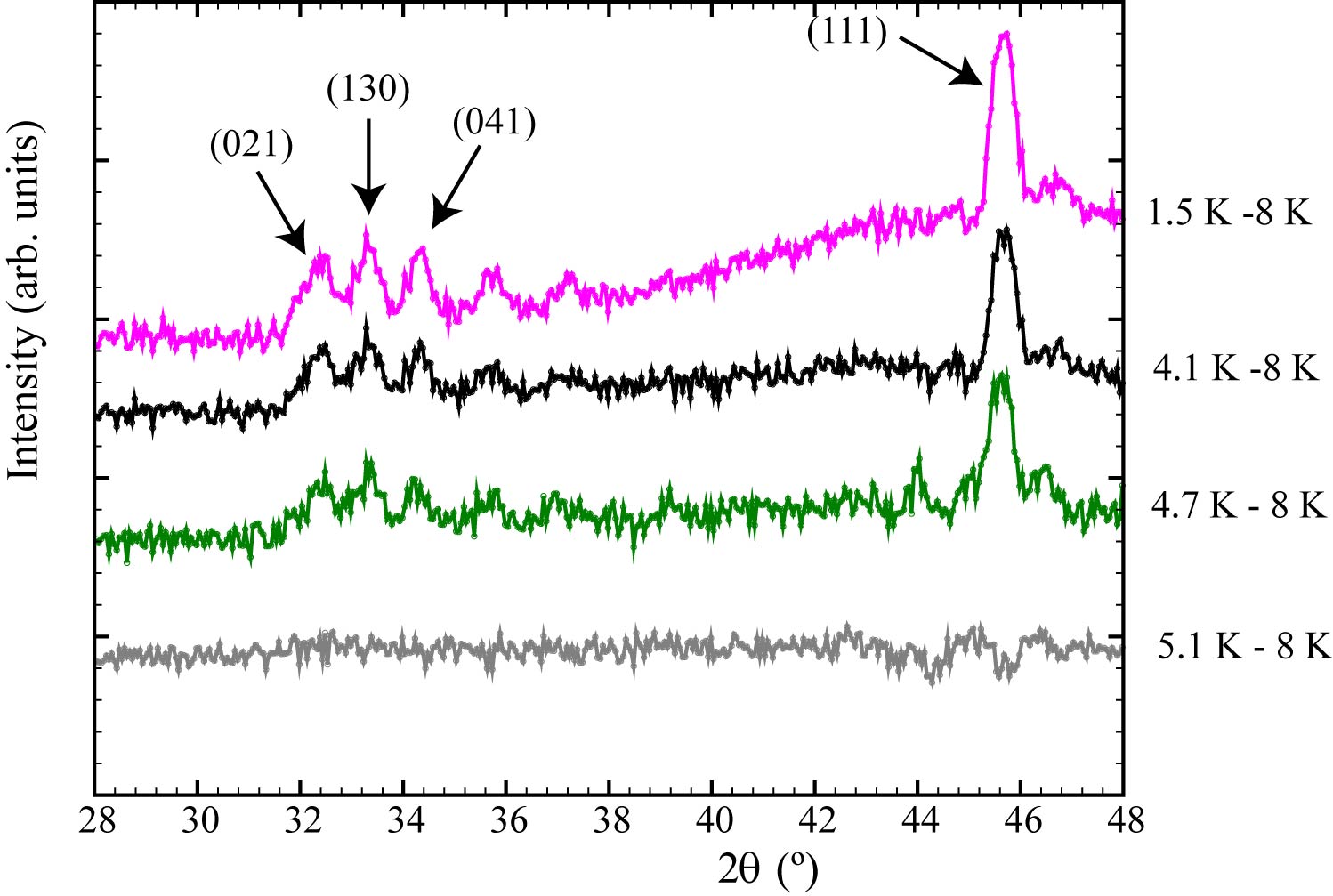}
		\caption{(a) Temperature-difference neutron powder diffraction patterns measured on the D20 diffractometer, obtained by subtracting the 8~K (paramagnetic-state) data from diffraction patterns collected at selected temperatures. The resulting spectra highlight the magnetic contribution and reveal magnetic Bragg reflections indexed as (021), (130), (041), and (111). The magnetic intensity decreases with increasing temperature and is reduced to a very weak residual signal at 5.1~K.}
		\label{NDS1}
	
\end{figure}

\begin{figure}

		\includegraphics[width=8cm, trim=4cm 3.5cm 4cm 4cm, clip]{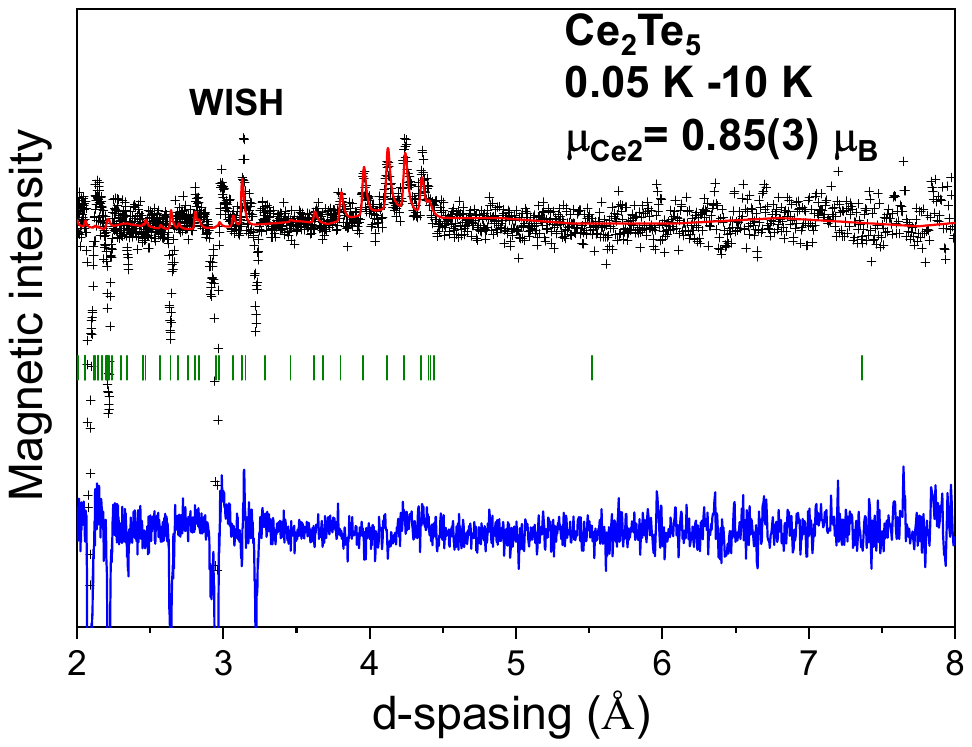}
		\caption{(a) NPD pattern of Ce$_2$Te$_5$ measured using the WISH time-of-flight  diffractometer and the data are plotted as temperature difference between the diffraction pattern at 0.05~K and that at 10~K, revealing the additional magnetic peaks intensity at 0.05~K, at the same positions at the nuclear Bragg peaks, due to the magnetic ordering of the Ce moments. The ordered moment size estimated from the refinement of the magnetic diffraction pattern  is 0.85(3)~$\mu_{\mathrm{B}}$ at 0.05~K on the Ce2 site, while the Ce1 site remains non-magnetic. The large positive and negative intensities (on the left) are due to thermal expansion of the sample and sample environment (subtraction of large Bragg peaks is not perfect). }
		\label{NDS2}
	
\end{figure}

Here we have presented the results of neutron diffraction experiments carried out using a dilution fridge on D20 at ILL (Fig.~\ref{NDS1}) and WISH at ISIS Neutron and Muon Source (Fig.~\ref{NDS2}).

For completeness, the structural parameters obtained from the Rietveld refinement of the room-temperature D2B NPD data are provided below. The refined lattice parameters are listed in Table~\ref{TS1}, and the corresponding atomic coordinates are given in Table~\ref{TS2}. The refinement confirms the orthorhombic $Cmcm$ crystal structure of Ce$_2$Te$_5$.

\begin{table}[ht]
\caption{Refined lattice parameters of Ce$_2$Te$_5$ obtained from the room-temperature D2B neutron powder diffraction data. The crystal structure was refined in the orthorhombic space group $Cmcm$ (No. ~63).}
\label{TS1}
\vskip 0.5cm
	\addtolength{\tabcolsep}{+10.0pt}
\centering
\begin{tabular}{cc}
\hline\hline
Parameter & Value \\
\hline
$a$ (\AA) & 4.42940(14) \\
$b$ (\AA) & 44.29740(149) \\
$c$ (\AA) & 4.44298(14) \\
$\alpha$ ($^\circ$) & 90 \\
$\beta$ ($^\circ$) & 90 \\
$\gamma$ ($^\circ$) & 90 \\
\hline\hline
\end{tabular}
\end{table}

\begin{table}[ht]
\caption{Refined atomic coordinates for Ce$_2$Te$_5$ obtained from the room-temperature D2B neutron powder diffraction data. All atoms occupy the Wyckoff position 4$c$ of the space group $Cmcm$ (No.~63).}
\label{TS2}
\centering
\vskip 0.4cm
	\addtolength{\tabcolsep}{+9.0pt}
\begin{tabular}{ccccc}
\hline\hline
Atom & Wyckoff & $x$ & $y$ & $z$ \\
\hline
Ce1 & 4$c$ & 0 & 0.09910(21) & 0.25000 \\
Ce2 & 4$c$ & 0 & 0.30676(20) & 0.25000 \\
Te1 & 4$c$ & 0 & 0.17145(27) & 0.25000 \\
Te2 & 4$c$ & 0 & 0.38138(22) & 0.25000 \\
Te3 & 4$c$ & 0 & 0.54053(24) & 0.25000 \\
Te4 & 4$c$ & 0 & 0.75001(27) & 0.25000 \\
Te5 & 4$c$ & 0 & 0.95915(28) & 0.25000 \\
\hline\hline
\end{tabular}
\end{table}

\section{Supplementary Inelastic Neutron Scattering Data}

\begin{figure*}
\includegraphics[width=14 cm, keepaspectratio]{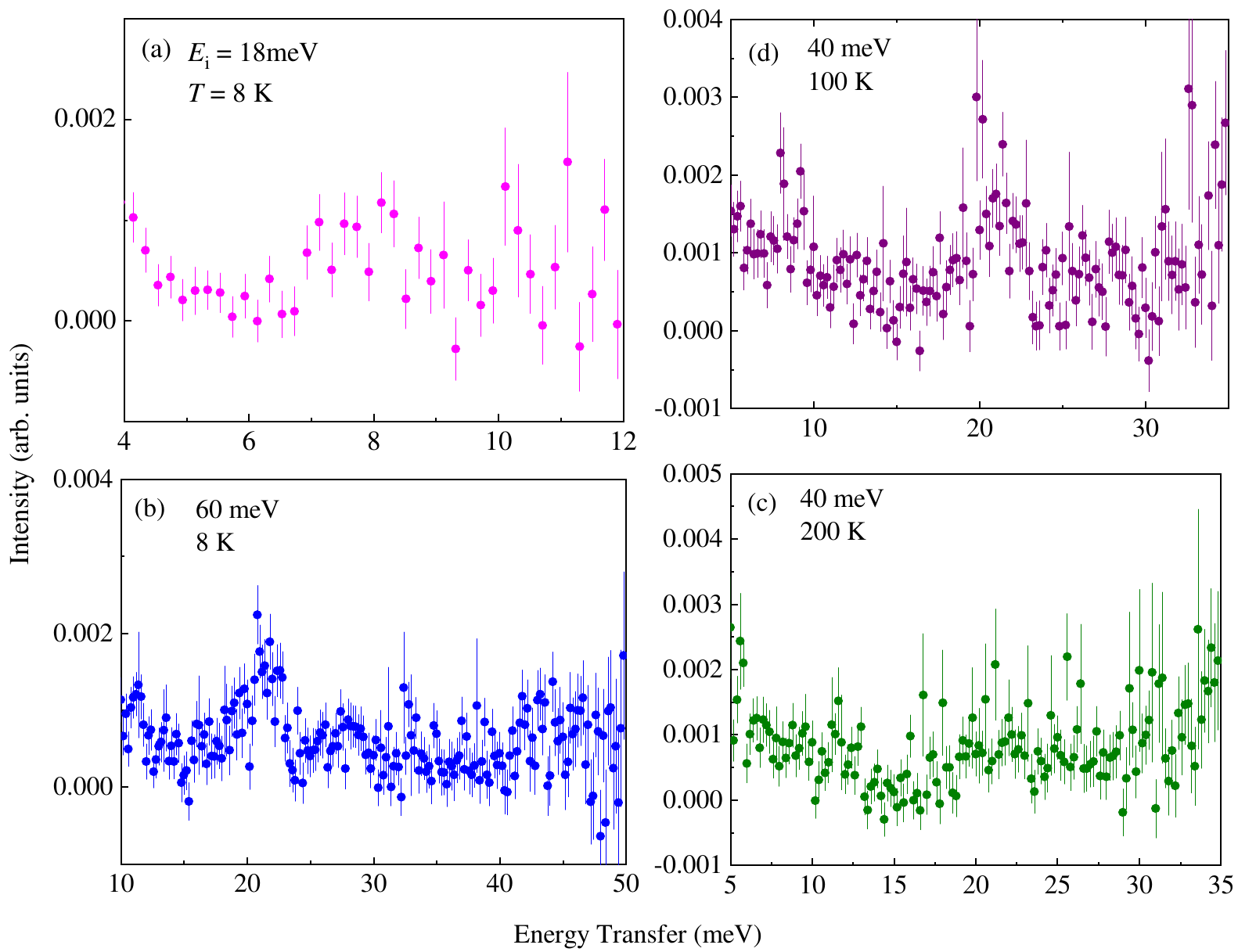}
\caption{
One-dimensional energy cuts of the INS spectra of Ce$_2$Te$_5$ at different incident neutron energies and temperatures. (a) $E_i = 18$~meV at $T = 8$~K,(b) $E_i = 60$~meV at $T = 8$~K, (c) $E_i = 40$~meV at $T = 200$~K, and (d) $E_i = 40$~meV at $T = 100$~K.}
\label{INS3}
\end{figure*}

To complement the main INS results discussed in the main text, Figs.~\ref{INS3}(a)–\ref{INS3}(b) present additional one-dimensional energy cuts measured at $T = 8$~K using different incident neutron energies. For $E_i = 18$~meV, a weak low-energy excitation centered near $\sim 8$~meV is consistent with the lowest CEF excitation identified in the $E_i = 40$~meV data [Fig.~\ref{INS2}(b)]. Moreover, the $E_i = 60$~meV spectrum reveals spectral weight at both $\sim 8$~meV and $\sim 21$~meV, corresponding to the two CEF excitations resolved in the 40~meV data along with a broad hump-like feature around $\sim 27$~meV. At present, it remains unclear whether this feature represents a genuine magnetic excitation, possibly associated with a second crystallographically inequivalent Ce site, or arises from residual phonon scattering, multiple scattering processes, or background effects. Given the weak magnetic signal and the absence of a nonmagnetic reference compound, a definitive assignment of this feature is not possible based on the current data. 

Figures~\ref{INS3}(c)-(d) show the INS spectra collected with $E_i = 40$ meV at $T = 200$ and 100~K, respectively. At these elevated temperatures, the magnetic signal is strongly suppressed, as expected, and dominated by phonon scattering and background contributions, resulting in a reduced signal-to-noise ratio. Consequently, these data do not allow a reliable extraction of CEF excitations but serve to demonstrate the magnetic nature of the 8 and 21 meV excitations based on the temperature evolution.

\bibliography{bibliography}

\end{document}